\pdfoutput=1
\documentclass[10pt, amsmath]{revtex4}
\usepackage{graphicx}
\usepackage{bm}
\usepackage{subfigure} 

\begin{document}
\title{Computation of nucleation of a non-equilbrum first-order phase transition using a rare-event algorithm}

\author{David A. Adams$^a$}%
 \email{davidada@umich.edu}
 \author{Robert M. Ziff$^b$}%
 \email{rziff@umich.edu}
 \author{Leonard M. Sander$^a$}%
 \email{lsander@umich.edu}
\affiliation{%
$^a$Department of Physics and $^b$Department of Chemical Engineering,
University of Michigan, Ann
Arbor MI 48109-2136.
}

\begin{abstract}
We introduce a new Forward-Flux Sampling in Time (FFST) algorithm to efficiently measure transition times in rare-event processes in non-equilibrium systems, and apply it to study the first-order (discontinuous) kinetic transition in the Ziff-Gulari-Barshad model of catalytic surface reaction.  The average time for the transition to take place, as well as both the spinodal and transition points, are clearly found by this method.
\end{abstract}

\maketitle

\section{Introduction}

In many systems, rare events occur with a very low probability compared to typical events. Sometimes they are of central interest.  Examples include the extinction of diseases \cite{andersson2000stochastic} or of populations \cite{bartlett1961stochastic}, network queue overflow \cite{medhi2003stochastic}, and slow chemical reactions \cite{van2007stochastic}. The study of such processes poses a particular challenge to simulation.  In the field of chemical physics, many rare-event techniques are commonly used: transition path sampling \cite{bolhuis2002t}, transition interface sampling \cite{van2003novel}, milestoning \cite{faradjian2004computing}, the string method \cite{W2005Finite}, and the weighted-ensemble method \cite{Huber1996Weighted}, to highlight a few methods. A thorough review can be found in \cite{dellago2008transition}. Most of these methods require that the system being studied has an underlying energy landscape, which precludes their use on non-equilibrium systems, i.e., systems that lack detailed balance. Forward Flux Sampling \cite{allen2005sampling} (FFS) is a rare-event technique designed specifically for non-equilibrium systems, and has proven useful in studying genetic switches \cite{allen2005sampling, allen2006simulating, valeriani2007computing}, nucleation \cite{valeriani2007computing, sanz2007evidence, allen2008homogeneous}, isomerization of alanine dipeptide \cite{Velez2009Kinetics}, and the Maier-Stein model of reaction dynamics \cite{maier1993effect, valeriani2007computing, allen2006forward}. In this paper, we develop a variant on FFS and use it to study the first-order non-equlibrium phase transition in a catalysis model.

FFS was developed to measure transition rates between two locally stable regions $A$ and $B$ separated by a high, featureless barrier.  When the barrier separating these regions contains long-lived metastable states, FFS is generally inaccurate or inefficient, depending on how it is applied.  We have overcome this limitation of FFS with a variant, which we call Forward Flux Sampling in Time (FFST).  In our method, we adjust for long-lived metastable states by measuring the times associated with sampling the region between $A$ and $B$ in the second stage of the FFS algorithm. The method is described in detail in Appendix A.

We apply FFST to the Ziff-Gulari-Barshad (ZGB) catalytic surface-reaction model \cite{ZiffGulariBarshad86}. This model is of interest because it has a first-order phase transition which acts in many ways like an equilibrium phase transition (it shows critical behavior, nucleation, etc.), but the model is manifestly non-equilibrium. Many techniques have been applied to the transition in order to tease out its properties.  FFST allows us to study the \textit{dynamics} instead of overall rates. Using FFST we found transition times for nucleation as large as $10^{40}$ Monte Carlo steps (MCS), more than $30$ orders of magnitude longer than those accessible to direct simulation. The method generates not only the transition times but the ensemble of most-likely states as the system progress from one phase to another.  This allows us to measure properties of the ensemble during the transition, which helps determine the pathway.

The outline of this paper is as follows:  In section \ref{sec:model} we describe the model and simulation method. In section \ref{sec:results} we describe our results, and in section  \ref{sec:conclusions} we summarize our conclusions.  In the Appendix A we give the details of the FFST technique, and in Appendix B we test FFST on an exactly soluble one-dimensional system.

\section{Model and Simulation Method}
\label{sec:model}
\subsection{The ZGB Model}
The ZGB model was introduced to study the behavior of the oxidation of carbon monoxide (CO) on platinum surfaces \cite{ZiffGulariBarshad86}.  The catalytic surface is represented by a square lattice on which CO and O$_2$ can adsorb. The CO takes up one lattice site whereas the O$_2$ dissociates into two O atoms which take two adjacent vacant sites. Both species, CO and O, are bound to the surface until the other species adsorbs on a neighboring site. At this point the CO and O form CO$_2$ and desorb from the catalyst leaving two lattice sites empty. The rates of reaction $\mathrm{CO} + \mathrm{O} \rightarrow \mathrm{CO}_2$ and the desorption of CO$_2$ are assumed to be infinite. The state of the catalyst is controlled by the fraction of the time CO is attempted to be placed on the catalyst; this fraction is called $p_\mathrm{CO}$.

In this basic model, there exists a region of steady-state reaction, bordered from below at $p_\mathrm{CO} = p_1$ by a second-order, continuous kinetic phase transition to an O-covered state, and above  $p_\mathrm{CO} = p_2$ by a discontinuous, first-order kinetic phase transition to a CO-covered state.  The first-order transition is robust to small changes in the model (diffusion, small desorption, etc.) and is seen experimentally at low temperatures as a sharp transition from high to low reactivity \cite{EhsasiEtAl89}.  The second-order O-poisoning transition is weak and not seen experimentally.  The first-order transition is associated with many complex oscillatory and wave phenomena in  theoretical \cite{MachadoBuendiaRikvold05,MachadoBueniaRikvoldZiff05} and experimental systems \cite{Jakubith..Ertl90, Kim..Ertl01} and has thus received much attention.  It also serves as a paradigm for general first-order kinetic phase transitions \cite{MarroDickman99}.

Studying the first-order transition has proved to be a  challenging problem in simulations. Because of the difficulty of nucleating a sufficiently large CO cluster or island, simply increasing $p_\mathrm{CO}$ from the reactive steady-state misses the transition point, and instead the CO-poisoning is seen to occur at $p_\mathrm{CO} \approx 0.5277$ \cite{MeakinScalapino87}. That point is close to an effective spinodal point $p^*$, where the transition occurs without any kind of barrier   \cite{BrosilowZiff92,LoscarAlbano09}.

\begin{figure}[h]
  \begin{minipage}[t]{.45\textwidth}
    \begin{center}
      \includegraphics[width=70mm]{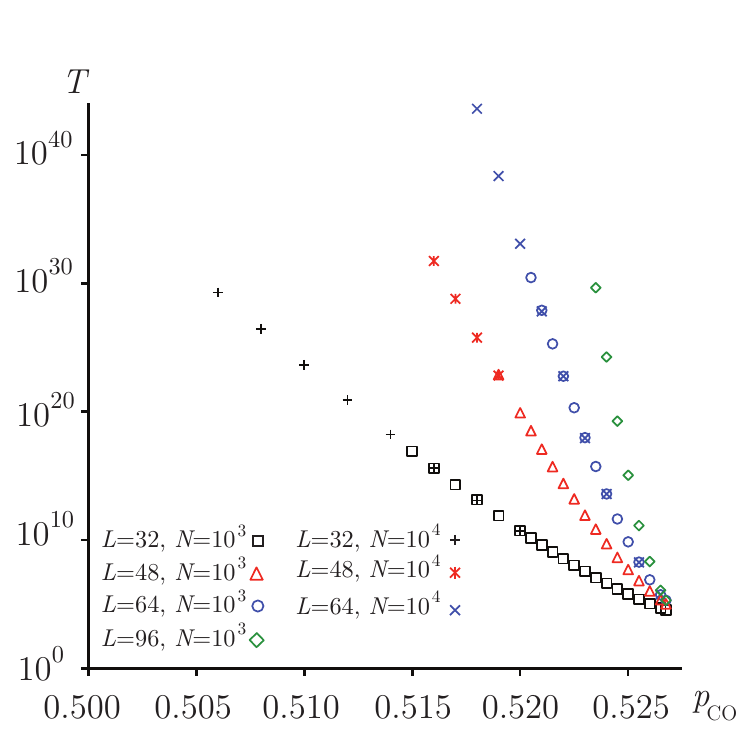}
      \caption{The poisoning time $T$ vs.\ $p_\mathrm{CO}$ for various values of system size $L$ and FFST trials $N$.}
      \label{fig:TransitionTime}
    \end{center}
  \end{minipage}
  \hfill
  \begin{minipage}[t]{.45\textwidth}
    \begin{center}
      \includegraphics[width=70mm]{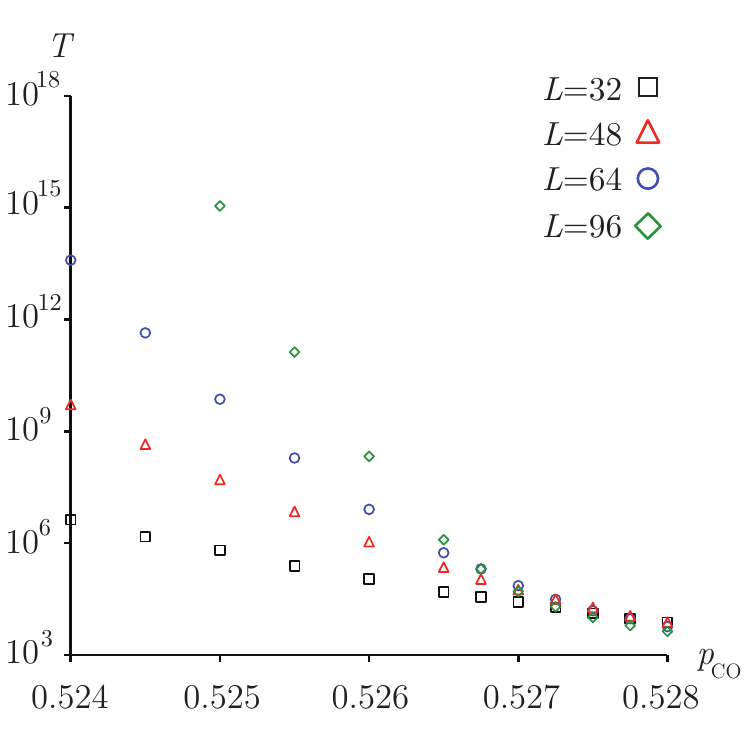}
      \caption{The poisoning time $T$ vs.\ $p_\mathrm{CO}$ for various values of $L$ with $N=10^3$. }
      \label{fig:TransitionTimeZoom}
    \end{center}
  \end{minipage}
  \hfill
\end{figure}

Estimating the value of $p_2$ and properties of the first-order phase transition have received significant attention. Using a ``constant-coverage" technique, the values $p_2 = 0.52560(1)$ \cite{ZiffBrosilow92,BrosilowZiff92} and more recently $0.525615(5)$ \cite{LoscarAlbano09} have been found. Other methods, including histograms \cite{TomeDickman93}, epidemic analysis \cite{EvansMiesch91}, and epidemics and trigger waves \cite{EvansRay94,GoodmanEtAl95}, have also been used to probe the first-order transition.

The results of these studies is that there is a first-order transition at $p_2 \approx 0.5256$, and a spinodal at $p^* \approx 0.527-0.528$   \cite{ZiffBrosilow92,LoscarAlbano09}, although the precise, and somewhat lower, value $p^* =  0.52675(5)$ has also been proposed   \cite{Albano01}. For $p_2 < p_\mathrm{CO} < p^*$, there presumably exists a critical CO-cluster size, below which clusters tend to shrink and above which they tend to grow.  That critical size changes from $\infty$ to something of the order of the lattice spacing as $p_\mathrm{CO}$ goes from $p_2$ to $p^*$.  For finite systems, the behavior is controlled strongly by the boundaries.  Using periodic boundary conditions, as $p_\mathrm{CO}$ is increased, the largest CO-cluster goes from being isolated, to wrapping around one direction (leading to two interfaces that are flat on the average) to wrapping around in both directions, as illustrated in Fig.~\ref{fig:ZGB_Picture}.  In the intermediate coverage region, the constant-coverage  method gives $p_2$ accurately with very small finite-size effects.

While the constant-coverage technique maps out the transition, it does not provide any information about the dynamics of the system.  For that, it is necessary to study the standard (constant-rate) ensemble.  But  in that case, the nucleation barrier makes it virtually impossible to study dynamics except for very close to the spinodal point.  To overcome this challenge, we use a modified FFS technique to find nucleation dynamics as well as overall rates for a wide range of $p_\mathrm{CO}$ values.

\subsection{The simulation method}

We study the ZGB model using the constant-rate ensemble on a square $L \times L$ lattice with periodic boundary conditions.  The dynamics involve repeated attempts to adsorb the CO or O$_2$ species. The procedure for an adsorption trial is given below.

\begin{itemize}
\item pick $r \in [0,1)$, if $r < p_\mathrm{CO}$, attempt to place a CO molecule, otherwise attempt to place O$_{\textrm{2}}$. \item pick a random lattice site $(x,y)$ and continue if the site is empty. If placing O$_{\textrm{2}}$, also pick a neighboring site $(x\pm 1, 0)$ or $(x, y \pm 1)$ and continue if that site is also empty. \item Place the CO or O$_{\textrm{2}}$ (dissociated) onto the empty lattice site(s). \item For each lattice site now occupied, check all of the neighbors of that site and determine if any of those neighbors are of the opposite species. If any are, remove a randomly chosen neighbor of the opposite species with the recently placed species.
\end{itemize}

\begin{figure}[t]
\includegraphics[width=70mm]{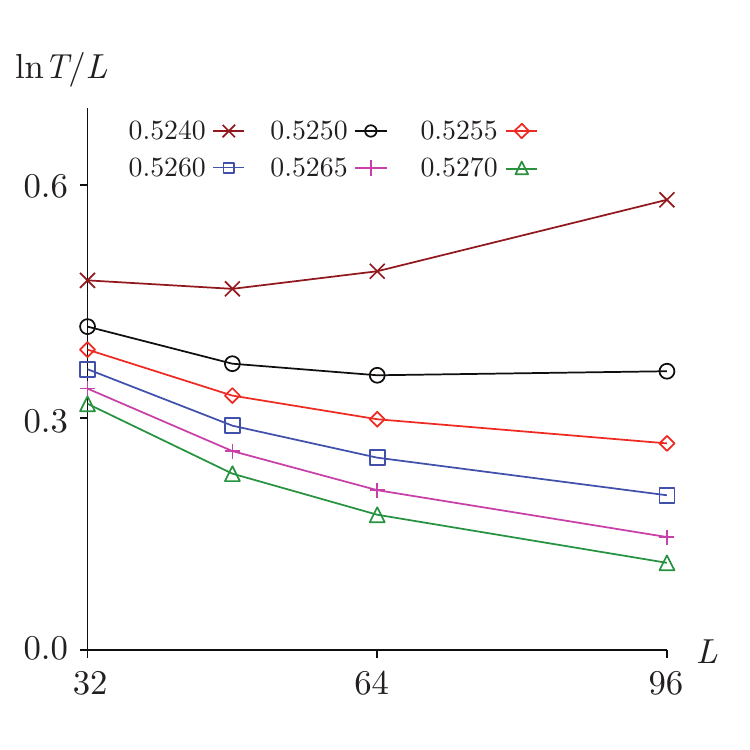}
\caption{\label{fig:ExitFit} The scaled transition time $\ln T / L$ vs.\ $L$ for various values of $p_\mathrm{CO}$.}
\end{figure}

Before any adsorption trials can be attempted, the system must be initialized. Because we wish to study the first-order phase transition, we prepare the system in the reactive state.  Starting from an initially empty lattice with $p_\mathrm{CO}$ close to the spinodal would frequently lead directly to the poisoned state. Instead, we prepare the system by adsorbing CO with probability $p^i_\mathrm{CO} = 0.07$ and O with probability $p^i_{\mathrm{O}} = 0.43$ on every site. This initialization generates invalid states with CO neighboring O.  To remedy this problem, we run the simulation for $10$ MCS, which drives the system to a valid state. These $10$ MCS ``burn off" a significant fraction of the original CO and O including those with neighbors of the opposite species.

\begin{figure}[h]
  \begin{minipage}[t]{.45\textwidth}
    \begin{center}
     \includegraphics[width=70mm]{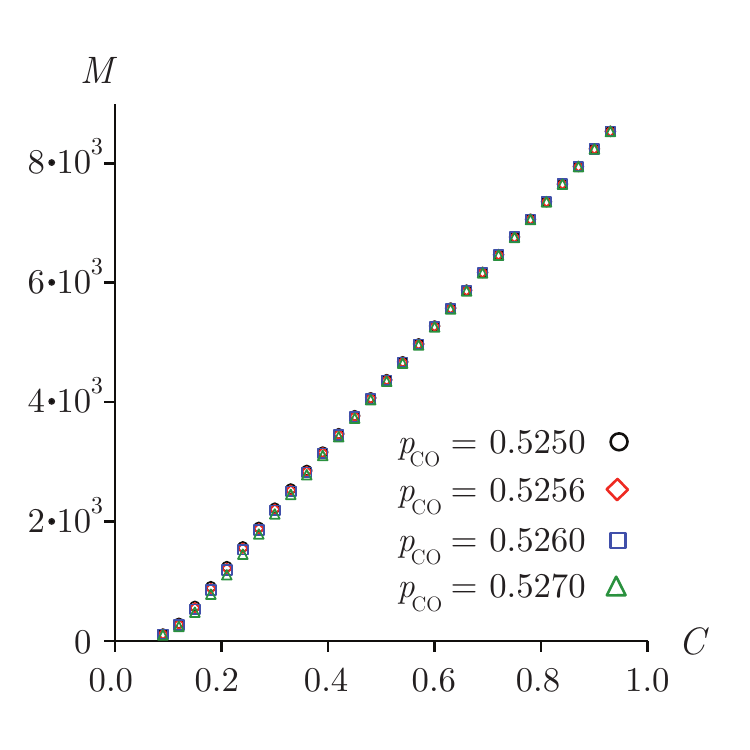}
      \caption{The average largest cluster size M vs.\ coverage $C$ for various values of $p_\mathrm{CO}$ and $L=96$.}
      \label{fig:LargestCluster_L96}
    \end{center}
  \end{minipage}
  \hfill
  \begin{minipage}[t]{.45\textwidth}
    \begin{center}
     \includegraphics[width=70mm]{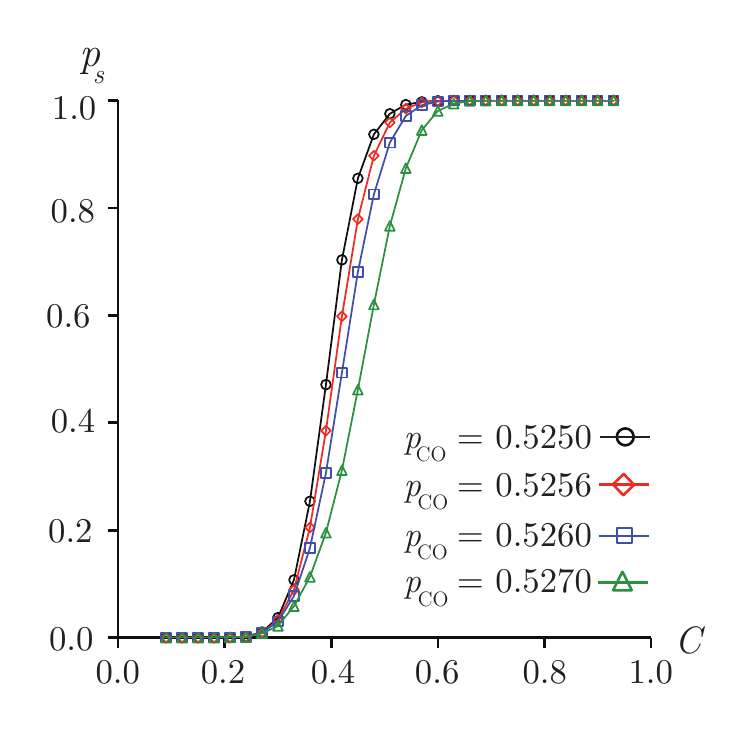}
      \caption{The spanning probability vs.\ $C$ for various values of $p_\mathrm{CO}$ and $L=96$.}
      \label{fig:SpanningCluster_L96}
    \end{center}
  \end{minipage}
  \hfill
\end{figure}

\subsection{Forward Flux Sampling in Time}

To study the first-order phase transition, we use the FFST algorithm, which is described in detail in Appendix A.  A sketch of the algorithm is as follows. Before the simulations begin, we define a starting region in state-space $A$ and an ending region $B$, bounded by `barriers' $\lambda_0$ and $\lambda_M$, respectively. We also define $\lambda_1$ through $\lambda_{M-1}$ as dividing surfaces in state space that effectively mark the distance between $A$ and $B$. The first step of the algorithm is to run a long simulation starting in $A$ and recording where the sample path crosses $\lambda_0$ going out of $A$, and the average time spent after crossing back into $A$ before leaving again; we call this the internal return time $T_{int}$. Next, we start sample paths along $\lambda_0$ where the initial simulation crossed, and continue them until they reach $\lambda_1$ or go back into $A$. The fraction of paths that reach $\lambda_1$ gives an estimate of the probability of reaching $\lambda_1$ without going back into $A$, $P(\lambda_1 | \lambda_0)$. We also keep track on the average time it took to reach $\lambda_1$ and go back into $A$ starting from $\lambda_1$. In the second step, we continue paths from the locations along $\lambda_1$, where the previous paths stopped, and run them until they reach $\lambda_2$ or go back inside $A$, crossing $\lambda_0$. The results give estimates for $P(\lambda_2 | \lambda_1)$ and the time it takes to reach $\lambda_2$ or $A$ from $\lambda_1$. This process is repeated, step by step, until $\lambda_M$ is reached on the $M^\mathrm{th}$ step. Finally, we use the results collected to calculate the overall transition time, which is given by the probability of reaching $\lambda_M$ from $\lambda_0$ without going into $A$ ($P(\lambda_M | \lambda_0)$) times the average time it takes to leave $A$. The overall transition probability is given by the product of the intermediate transition probabilities, $P(\lambda_M | \lambda_0) = \prod_{i=0}^{M-1} P(\lambda_{i+1} | \lambda_i)$. The time to leave $A$ is the average time it takes to return to $\lambda_0$ from inside $A$, $T_{int}$, plus the average time it takes to return to $A$ from outside $\lambda_0$, $T_{ext}$, which we calculate from the times measured during the second stage of the algorithm. The gains of FFST over FFS are illustrated in Appendix B.

Thus, the technique follows fruitful paths from the reactive state to the CO-poisoned state. To apply the FFST to the ZGB model, we must first define an order parameter, which is used to determine progress towards the poisoned state. This function should smoothly increase as system transitions from being reactive to poisoned. We chose the fraction of CO on the lattice ($C$) as our order parameter.

FFST uses the order parameter to make surfaces (barriers) in state-space which are used to mark progress. Simulations are run from locations along a barrier in state space until they reach the next barrier, as defined by the order parameter, or the first barrier.

These barriers can be placed before the sampling begins, which we call \textit{static} barriers. The disadvantage of static barriers is that without \textit{a priori} information about how to best place them, some barriers will have a large effective separation creating a performance bottleneck. To overcome these performance issues, two methods have recently been developed \cite{Borrero08, Adams2010Barrier} which determine where to place the barriers automatically. To measure the poisoning time, we use the dynamic barrier placement method from \cite{Adams2010Barrier} to place the barriers during the FFST algorithm. We use static barriers for measurements of the average largest cluster, largest cluster spanning probability, and committor probability \cite{bolhuis2002t}; these terms are defined below.  We use static barriers for the obserables because we want measurements at uniformly spaced intervals of coverage. When placing the barriers dynamically, we space them such that typically $10$\% of trials make it to the next barrier before returning to the steady-state of the reactive region with low CO coverage. We use $N$ trials per step in the FFST algorithm, except for the first barrier, which we use $10N$; we use $N/10$ trials to determine where the next barrier should be placed. We locate the first barrier at the largest value of the order parameter found in the first 50 MCS. For the calculation of poisoning times, we used $N=10^3$ and $10^4$. $N=10^4$ was used for average largest cluster, largest cluster spanning probability, and committor probability.

\begin{figure}[h]
  \begin{minipage}[t]{.45\textwidth}
    \begin{center}
      \includegraphics[width=70mm]{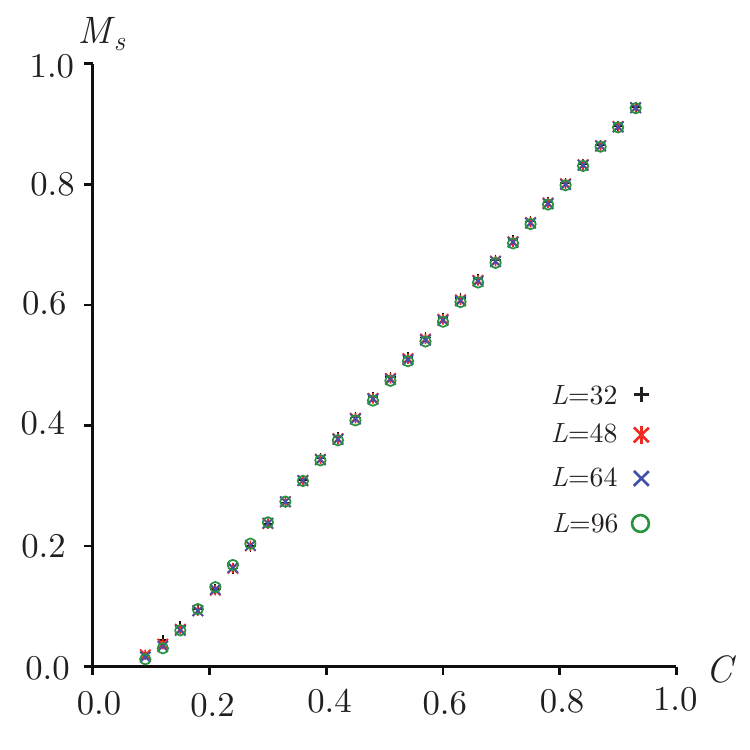}
      \caption{The average largest cluster size $M_s$ vs.\ coverage $C$,  as a function of $L$ for $p_\mathrm{CO}=0.5256$.}
      \label{fig:LargestCluster_pco5256}
    \end{center}
  \end{minipage}
  \hfill
  \begin{minipage}[t]{.45\textwidth}
    \begin{center}
      \includegraphics[width=70mm]{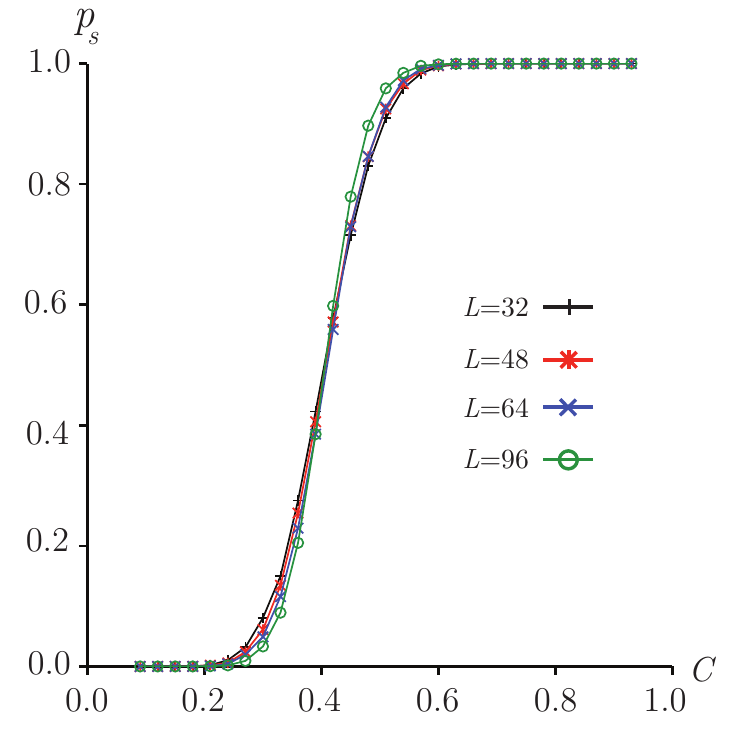}
      \caption{The spanning probability $p_s$ vs.\ coverage $C$ for various $L$ and $p_\mathrm{CO}=0.5256$.}
      \label{fig:SpanningCluster_pco5256}
    \end{center}
  \end{minipage}
  \hfill
\end{figure}

\section{Results}
\label{sec:results}

We desire the transition times from the reactive to poisoned state, which we call the \emph{poisoning time} $T$. We determine $T$ for a range of $L$: $32$, $48$, $64$, and $96$, as well a number of values of  $p_\mathrm{CO}$ in the critical region, $p_2 < p_\mathrm{CO} < p^*$, and in the reactive region, $p_\mathrm{CO} < p_2$ down to $p_\mathrm{CO} = 0.505$. The results are shown in Figs.~\ref{fig:TransitionTime} and \ref{fig:TransitionTimeZoom}.

Figs.\ \ref{fig:TransitionTime} and \ref{fig:TransitionTimeZoom} show that the poisoning time grows smoothly as $p_\mathrm{CO}$ decreases, with no indication of a transition at $p_2$.  Even if $p_\mathrm{CO}$ is much smaller than $p_2$, we found it possible to measure the poisoning time, demonstrating that the reactive state for finite systems is always metastable. We were able to determine maximum poisoning times of $T = 10^{30}$ - $10^{40}$ MCS.  The poisoning times converge to a value that is independent of system size for $p_\mathrm{CO} \approx \ 0.5275$, as shown in Fig.~\ref{fig:TransitionTimeZoom}.   We associate this point with the spinodal point $p^*
$: as it represents the spontaneous and simultaneous nucleation of multiple clusters throughout the system, and is not influenced by the boundaries of the system.

We attempt to find the form of the transition time as a function of $L$ and $p_\mathrm{CO}$. In many non-equilibrium systems, transition times  take the form $T \approx e^{W L}$, where $W$ is an effective energy barrier and $L$ characterizes the size of the system.  Fig.~\ref{fig:ExitFit} shows $\ln T/L$ which is an estimate for $W$. We see that $\ln T/L$ does not have a strong system size dependence near $p_2$, which may indicate that there is an effective Arrhenius energy. If this were the case, then $\ln T/L$ will become independent of $L$ for very large system sizes.

\begin{figure*}[t]
\includegraphics[width=160mm]{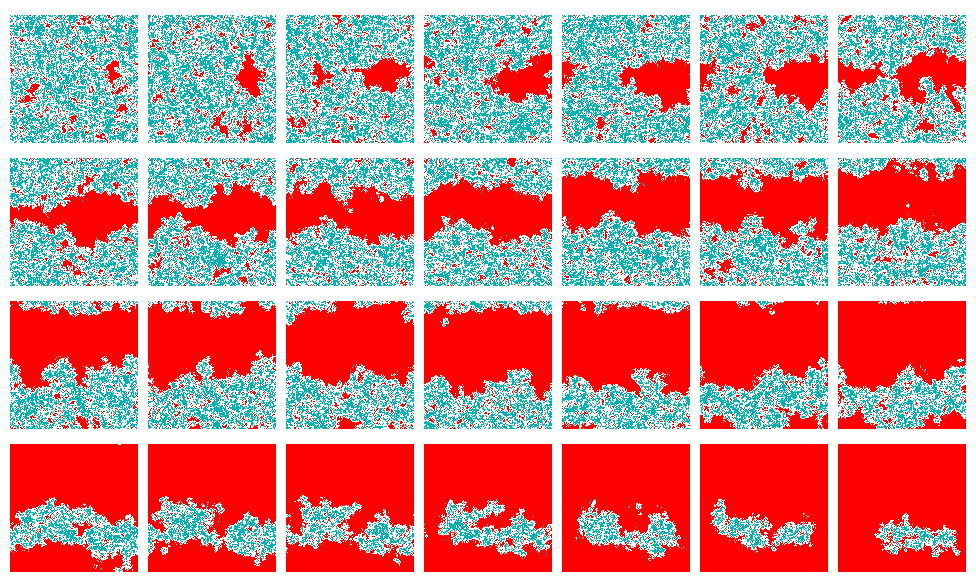}
\caption{\label{fig:ZGB_Picture} The evolution of the most likely path to the poisoned state for $L=128$ and $p_\mathrm{CO} = 0.5256 \approx p_2$. The pictures are in increasing values of coverage fraction from left to right starting at $0.09$ and increasing in steps of $0.03$.}
\end{figure*}

Although FFST is primarily a tool for obtaining transition times, it also gives ensembles of states along each barrier as it progresses towards poisoning. These ensembles at a fixed value of coverage represent essentially what the constant-coverage ensemble attempts to mimic. Expectation values of quantities, like largest cluster size, can be taken over these ensembles, which give insight into the dynamics of the phase transition.

We ran FFST with evenly spaced static barriers and measured the average largest cluster size and the spanning probability for the ensemble captured on every barrier. Spanning occurs when the largest cluster wraps around the periodic boundary and touches itself. We see a dependence on $p_\mathrm{CO}$ and $L$ upon the spanning probability, as shown in Figs.~\ref{fig:SpanningCluster_L96} and \ref{fig:SpanningCluster_pco5256}.  Fig.~\ref{fig:SpanningCluster_L96} shows that the smaller the value of $p_\mathrm{CO}$, the more likely that the largest cluster will wrap earlier in the path to poisoning, at fixed $L$. Fig.~\ref{fig:SpanningCluster_pco5256} shows that increasing $L$ appears to narrow the range of coverage for which spanning has a non-negligible probability of occurring or not occurring, at fixed $p_\mathrm{CO} \approx p_2$. This suggests that the variation in shape of the largest cluster decreases with $L$.  It also appears that reaching a spanning probability of $50$\% is achieved at $C \approx 0.42$ independent of $L$ at $p_2$. This indicates that the average largest cluster is significantly elongated when it begins to span the system; as seen in Fig.~\ref{fig:ZGB_Picture}. We also found a linear relationship between the scaled average largest cluster size and coverage, independent of both $L$ and $p_\mathrm{CO}$, as shown in Figs.~\ref{fig:LargestCluster_L96} and \ref{fig:LargestCluster_pco5256}.

\begin{figure}[t]
\includegraphics[width=120mm]{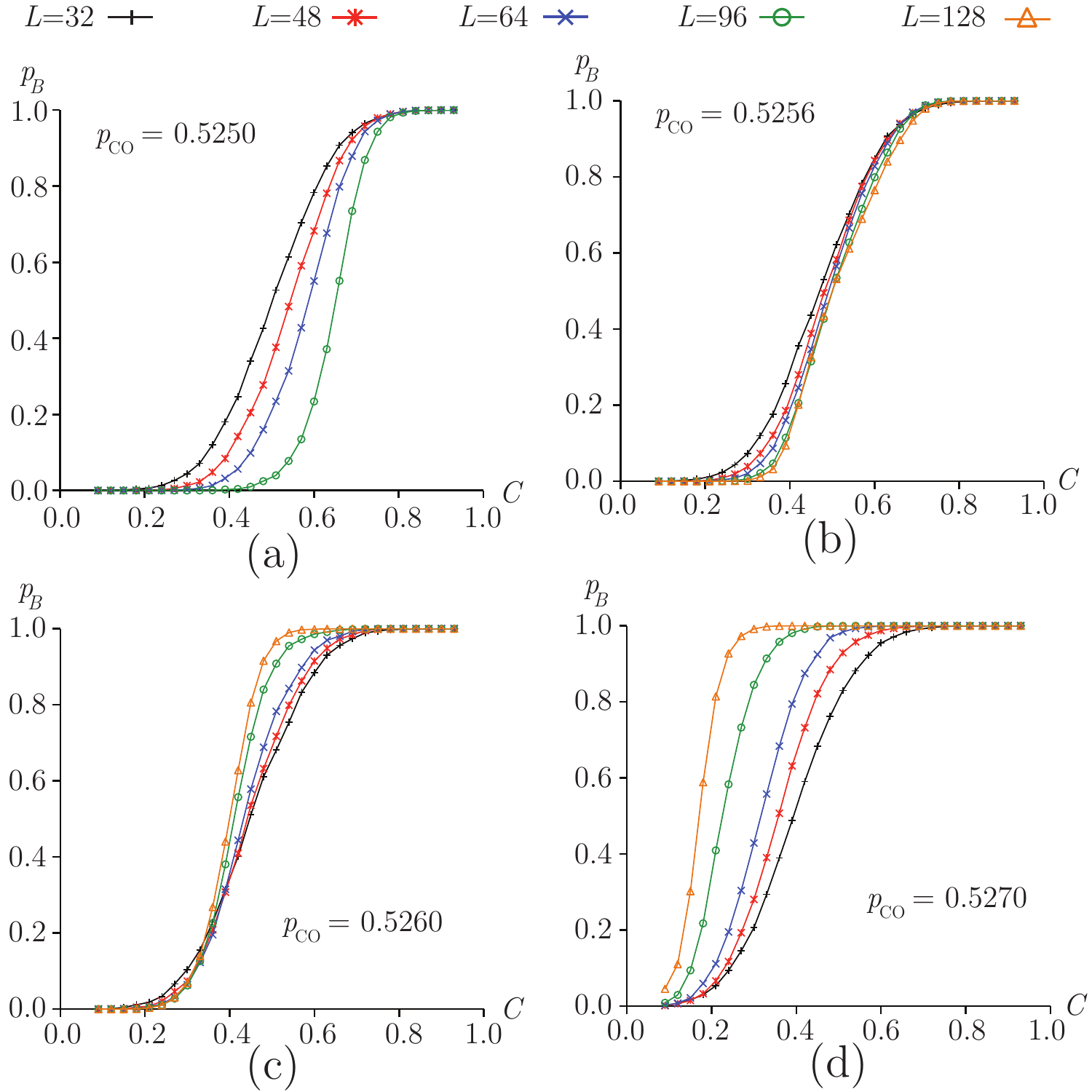}
\caption{\label{fig:Comm} The committor probability $p_B$ vs.\ $C$ for four values of $p_\mathrm{CO}$ and different $L$. $N=10^4$ for all data shown. }
\end{figure}

The ensembles of states at different values of the coverage can also be used to directly measure progress towards poisoning. By running every state in the ensemble until it returns to the reactive ($A$) or poisoned ($B$) state, one obtains the probability of poisoning from these particular values of the coverage. This probability as a function of the order parameter is called the committor probability, $p_B$. (In the case of the ZGB model, this is only an estimate because the model is non-equilibrium, so the forward-tending ensemble obtained is not necessarily the same as the steady-state non-equilibrium distribution along the barriers.) We measured the committor probability for various $L$ and $p_\mathrm{CO}$, as shown in Fig.~\ref{fig:Comm}. We find that for $p_\mathrm{CO}$ below $p_2$, the larger the system, the larger the coverage must be in order to have a particular probability of poisoning. The opposite effect is found above $p_2$. At $p_2$, the committor probability tends to a single form for the largest system sizes. This shows the significance of $p_2$, as the point where $p_B$ is independent of $L$.

By keeping track of which state on a given barrier is responsible for a particular state on the next barrier, we were able to piece together complete trajectories from the reactive state to the poisoned state. Among these reconstructed paths it is possible to determine a most likely path. Every state $j_i$ on every barrier $\lambda_i$ has some probability of continuing to the next barrier before returning to the reactive state denoted by $P^j( \lambda_{i+1} | \lambda_i )$, which is measured during FFST. We define the most likely path as the path which connects $\lambda_0$ and $\lambda_M$, which has the largest value of the product of the intermediate barrier crossing probabilities, $\prod_{i}  P^j (\lambda_{i+1} | \lambda_i)$. The states in Fig.~\ref{fig:ZGB_Picture} show the most likely path for $L=128$ at the transition point $p_2$. The preferred path  to poisoning involves wrapping around the system, then expanding to complete CO coverage. This path is favored because nucleating droplets have an effective kinetic surface tension which causes droplets to be unfavored. Once a droplet spans the system and the net curvature disappears, the cluster is significantly more favored. This preferred pathway of wrapping and expanding has also been seen in magnetic memory switching \cite{weinan2003energy}.  This behavior is to be expected, at least near $C \approx 0.5$, where the cluster usually wraps around in one direction. In that case, at the transition point $p_2$, the system should be equally likely to poison or return to the reactive state, so $p_B \approx 0.5$ independent of $L$.

\section{Conclusions}
\label{sec:conclusions}

In this article, we introduce forward flux sampling in time (FFST). We use it to analyze the first-order phase transition in the ZGB model. We found a size-independent poisoning time at $p^*\approx 0.5275$, which is associated with the spinodal point. The poisoning time is a continuous function of $p_\mathrm{CO}$ near the first-order transition.  By inspecting the ensembles of states measured at each barrier, we found a linear relationship between the scaled average largest cluster size and the total CO coverage, which is practically independent of $L$ and $p_\mathrm{CO}$. When the paths from barrier to barrier are connected, they make the ensemble of successful trajectories. We found the most probable of these trajectories and found that wrapping and then expanding is the preferred path to poisoning.

At $p_\mathrm{CO} = 0.5256$, the committor probabilities appears to be independent of system size for large $L$. We believe that this is a signature of a first-order transition for the following reasons: first, the committor probability is $0.5$ for half coverage, which is has been previously used to determine the transition point \cite{ZiffGulariBarshad86}.  Second, the matching of the entire curves for various $L$ stems from the critical cluster size being infinite at the transition point. This implies that the probability of a droplet growing is always less than $1/2$. The growth probability cannot be strongly dependent on the cluster size as clusters of arbitrarily large size must all have roughly the same growth probability, slightly less than $1/2$. For large lattice sizes,  what matters is how close the cluster is to spanning, which is only dependent on the mass of the largest cluster scaled by the total system size. Thus, we have found that the committor probability can be used to locate the first-order transition point of the system. Lastly, we found size-dependent poisoning times for systems well below the transition point.

For the ZGB model, we found that the efficient implementation of FFS, i.e., a fixed number of crossings in the first stage, gives accurate results. This shows that ZGB doesn't have extremely long-lived metastable states, which is the primary advantage of FFST. But, we still found that FFST can outperform FFS in terms of smaller variance, which translates into better computational efficiency to reach a target variance.  Specifically, we found that FFST was effectively $35$\% more efficient than FFS for a test case.

\section{Acknowledgments}

This research was supported in part by the National Science Foundation through TeraGrid resources \cite{catlett2007teragrid} provided by Purdue University under grant TG-PHY090106 and through DMS-0553487. D. Adams would like to thank C. Fink for useful conversations.

\begin{figure}[t]
\includegraphics[width=140mm]{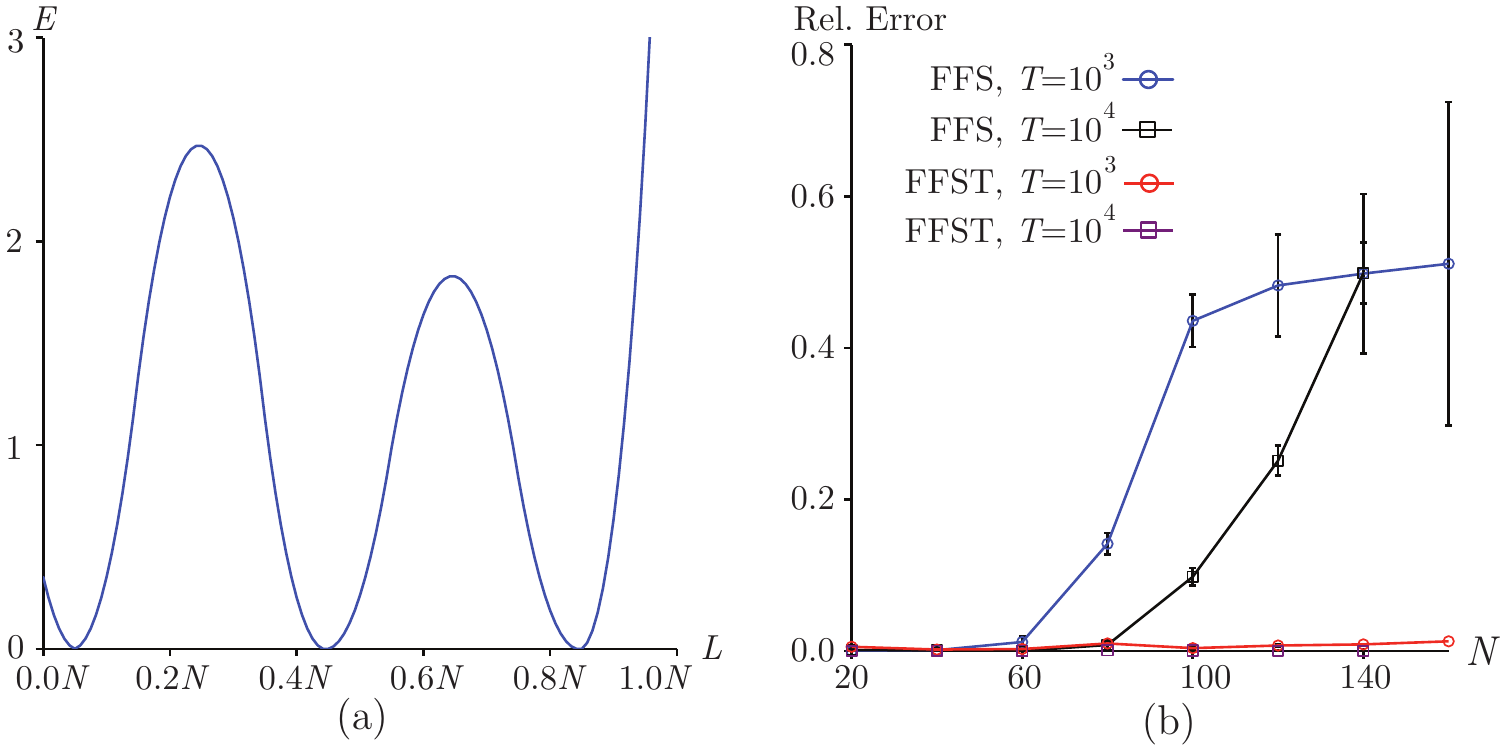}
\caption{\label{fig:Accuracy} (a) A sketch of the effective energy landscape as a function of scaled system length $L$ for a model 1D problem. (b) The relative error of the estimated transition time as a function of system length $N$.}
\end{figure}

\section{Appendix A --- The FFST algorithm}
\label{sec:AppendixA}

The FFS algorithm works well for problems with a featureless barrier, i.e., problems with infrequent but fast transitions. But, in the case of problems with \textit{slow} transitions, typically caused by long-lived metastable states, it has significant shortcomings \cite{allen2009forward}.  In these cases, FFS either grossly underestimates the transition time or becomes nearly as inefficient as direct simulation (see Appendix B for an example). In this case FFS produces a constant gain in simulation efficiency, of the order of ten times faster, whereas the `efficient' approach, where a fixed number of crossings is used to calculate the flux rate, can be $10^{30}$ times faster, as we found in this paper. In this Appendix, we derive the version of FFS, which we call FFST. We then outline two other variants on FFS which helps us illustrate the connection between FFS and the barrier method which we recently introduced \cite{Adams2010Barrier}.

As we explained above, FFS separates the problem of finding rare transition events into two problems: First, finding the rate of leaving the initial region $A$, and second, finding the probability of reaching the final region $B$ from the surface of $A$ without going back into $A$. Finding the rate of leaving $A$ typically involves running single a long simulation until it has exited $A$ a fixed number of times, say $10 N$ where $N$ is the number of trials per barrier in the second step. The number of crossings divided by the total time of the simulation (discounting paths that reach $B$) gives an estimate of the rate of leaving $A$. This calculation of the flux is accurate only if the trajectories sample capture the important times and features of the entire landscape. For example, if none of the trajectories generated sample a long lived, metastable state between $A$ and $B$, the estimated flux would be higher than the true flux. We show such an example in Appendix B.

A simple and correct way to avoid this problem is to run the initial simulation until it reaches $B$ one or more times. However, this is impractical because it effectively solves the problem using brute force, which is what FFS was designed to avoid, and would lead to miniscule efficiency gains over brute-force simulations. Our version of FFS has the accuracy of the `correct' calculation and the efficiency of the usual flux calculation.

\subsection{Forward Flux Sampling in Time}

Forward flux sampling in time (FFST), the algorithm we use in this paper, performs FFS in terms of transition times instead of rates. The problem is decomposed in the same way as FFS\cite{allen2005sampling}, utilizing the idea of endpoint regions $A$ and $B$. In this view, there are three important times: the time to return inside of $A$ from the surface of $A$ ($T_{ext}$), the time to get back to surface of $A$ from just inside $A$ ($T_\mathit{int}$), and the time to reach $B$ from the surface of $A$ ($T_f$) without going back into $A$. The only additional quantity needed to calculate the transition time, $T_{tot}$, is the probability $p$ of reaching $B$ from the surface of $A$. Then:
\begin{equation}
\label{eq:TrueTime}
T_{tot} = \left(\frac{1}{p} - 1\right)(T_{ext}+T_{int}) +  T_f.
\end{equation}
If the first term, which represents the total time spent on unsuccessful attempts to reach $B$, is much larger than the time of a successful attempt to reach $B$, which is explicitly used in the construction of FFS, $[ (1/p)-1](T_{ext}+T_{int}) \gg T_f$, and $p$, is small then the above formula simplifies to the inverse of the FFS formula,
\begin{equation}
\label{eq:TrueTimeFFS}
T_{tot} \approx \frac{1}{p} (T_{ext} + T_{int})  = \frac{1}{P(\lambda_{M} | \lambda_{0})} \left( \frac{h_A}{\Phi_{A,0}} \right) = \frac{1}{k_{AB}}.
\end{equation}

The efficiency gain of FFST comes from measuring $T_\mathit{int}$ during the first step, and measuring $T_\mathit{ext}$ and $p$ in the second step. Measuring $T_\mathit{int}$ involves counting the time that the initial simulation spent in region $A$.  In the second step, paths are allowed to run from the $i^{\textrm{th}}$ barrier to the  $(i+1)^{\textrm{th}}$ or back to $A$. We record the times to make these transitions to the next barrier ($t_i^{i+1}$) and back into $A$ ($t_i^0$) in addition to the probability of making it to the next barrier $P(\lambda_{i+1} | \lambda_{i})$. $T_{ext}$ can be calculated by properly weighting the time it takes for different paths to return to $A$,
\begin{equation}
\label{eq:Texternal}
\begin{split}
T_\mathit{ext} &= \sum_{i=0}^{M-1} P(\lambda_i | \lambda_0) P(\lambda_0 | \lambda_i) \left( t_0^i + t_i^0 \right) / \left( 1 - P(\lambda_M | \lambda_0) \right)  \\
&= \left[ \sum_{i=0}^{M-1}  \left( \prod_{j=0}^{i-1} P(\lambda_{j+1} | \lambda_j) \right)  (1- P(\lambda_{i+1} | \lambda_{i}) )  \left( \sum_{k=0}^{i-1} t_k^{k+1} + t_i^0  \right) \right] \bigg{ / }  \left( 1 - \prod_{i=0}^{M-1} P(\lambda_{i+1} | \lambda_{i}) \right).
\end{split}
\end{equation}
The numerator in the last expression is the sum over return times of all paths sorted by largest excursion. The probability that a path will start at $\lambda_0$, make it to $\lambda_i$ and then return to $\lambda_0$ without making any further progress is product of the probability of reaching $\lambda_i$ ($\prod_{j=0}^{i-1} P(\lambda_{j+1} | \lambda_j)$) and the probability of then returning to $\lambda_0$ without reaching $\lambda_{i+1}$ ($1- P(\lambda_{i+1} | \lambda_{i})$). The average time of this excursion is the time it takes to reach $\lambda_i$ ($\sum_{k=0}^{i-1} t_k^{k+1}$) plus the time it takes to return to $\lambda_0$ without making any more progress ($t_i^0$). The term in the denominator is for normalization and represents the probability of returning to $A$.  The expression for $T_f$ is much simpler: $T_{f} = \sum_{i=0}^{M-1} t_i^{i+1}$. These values of $T_\mathit{int}$, $T_\mathit{ext}$, and $T_f$ can be combined with $P(\lambda_M | \lambda_0)$ in (\ref{eq:TrueTime}) to obtain $T_\mathit{tot}$. By measuring $T_{ext}$ in the second step of FFS, the value of $T_{tot}$ obtained will be comparatively more accurate and have less variance. The only extra work done in FFST over FFS is keeping track of the simulation time during the second step, which makes the advantages gained by FFST practically free.

\subsection{Forward Flux with Quasi-Markov Dynamics}

In order to gain more insight into the connection between the FFS and the barrier method, we formulate FFS as quasi-Markov dynamics. That is, we iteratively calculate the time it takes to travel between three barriers. The first step is similar to FFS: run a single long simulation and calculate the time it takes to reach the first barrier ($\lambda_1$) from the surface of $A$ ($\lambda_0$), and where along $\lambda_0$ the sample crosses going out. $N$ samples are started along $\lambda_1$, where the initial path crossed in the first step, and are run until they reach $\lambda_2$ or $\lambda_0$. The average times to go from $\lambda_0$ to $\lambda_1$, $\lambda_1$ to $\lambda_2$, and from $\lambda_1$ to $\lambda_0$ are $t_0^1$, $t_1^2$, and $t_1^0$, respectively. In general, the time takes to reach $\lambda_j$ from $\lambda_i$ is given by $t_i^j$. The probability of reaching $\lambda_2$ from $\lambda_1$ without first going back to $\lambda_0$ is $P(\lambda_2 | \lambda_1)$.  These times and probabilities can be used to make a random walk with three states, $\lambda_0$, $\lambda_1$, and $\lambda_2$. The transition time from $\lambda_0$ to $\lambda_2$ ($t_0^2$) is given by the weighted times of all possible paths from $\lambda_0$ to $\lambda_2$. These paths can be organized by the number of times they return to $\lambda_0$. Writing out the first few terms in this series exposes the general form,
\begin{equation}
\label{eq:QuasiMarkov}
\begin{split}
t_0^2 &=  \left[ (t_0^1 + t_1^2) P(\lambda_2 | \lambda_1) \right] + \left[(t_0^1 + t_1^2 + (t_1^0+t_0^1)) P(\lambda_2 | \lambda_1) (1 - P(\lambda_2 | \lambda_1)) \right]  \\
&+  \left[(t_0^1 + t_1^2 + 2(t_1^0+t_0^1)) P(\lambda_2 | \lambda_1) (1 - P(\lambda_2 | \lambda_1))^2 \right]  +   \ldots  \\
&= \sum_{k=0}^\infty (t_0^1 + t_1^2 + k(t_1^0+t_0^1)) P(\lambda_2 | \lambda_1) (1 - P(\lambda_2 | \lambda_1))^k.
\end{split}
\end{equation}
The above expression is the sum over all possible ways to reach $\lambda_2$ from $\lambda_0$, sorted by the number of times the simulation returned to $\lambda_0$. The probability of returning to $\lambda_0$ $k$ times before reaching $\lambda_2$ is given by the product of the probability of not reaching $\lambda_2$ $k$ times ($(1 - P(\lambda_2 | \lambda_1))^k$) and then reaching $\lambda_2$ on the $(k+1)^\mathrm{th}$ try ($P(\lambda_2 | \lambda_1)$). The time this takes is given by the sum of the time it takes to go from $\lambda_0$ to $\lambda_1$ to $\lambda_0$ $k$ times ($k(t_1^0+t_0^1)$) plus the time it takes to make it from $\lambda_0$ directly to $\lambda_2$ ($t_0^1 + t_1^2$). Note that in practice the sum converges quickly because of the factor, $(1 - P(\lambda_2 | \lambda_1))^k$.

The next step is to repeat the same process using $\lambda_0$, $\lambda_2$, and $\lambda_3$ as the three barriers. Brute-force dynamics are used to measure $t_2^3$, $t_2^0$, and $P(\lambda_3 | \lambda_2)$.  Then, the three-barrier calculation from (\ref{eq:QuasiMarkov}) is used with $0$, $2$, and $3$, in place of $0$, $1$, and $2$. The result from the calculation is an estimated value of $t_0^3$. In general to calculate $t_0^{i+1}$ for the $i^{\textrm{th}}$ step we use:
\begin{equation}
\label{eq:QuasiMarkovGeneral}
t_0^{i+1}= \sum_{k=0}^\infty \left[ t_0^i + t_i^{i+1} + k(t_i^0+t_0^i) \right] P(\lambda_{i+1} | \lambda_i) (1 - P(\lambda_{i+1} | \lambda_i))^k.
\end{equation}
This process of performing short brute-force simulations, followed by solving (\ref{eq:QuasiMarkovGeneral}), is repeated for every barrier until $\lambda_M$ is reached. At the end we have $t_0^M = T_{tot}$. This method could be useful for practical simulations. Here we introduce it as a pedagogical device to show that by making small changes to the FFS algorithm the barrier method can be effectively obtained.

\subsection{Forward Flux Barriers}
We can look at quasi-Markov dynamics in another way.  We start by measuring the average time it takes to reach $\lambda_1$ starting at $\lambda_0$ ($t_0^1$), while also keeping track of where along $\lambda_1$ the path crosses going out. The paths are continued from $\lambda_1$ until they reach $\lambda_2$ or $\lambda_0$. If a sample reaches $\lambda_0$ then it is restarted at $\lambda_1$ at one of the locations where paths ended in the first step, and $t_0^1$ is added to the time. This process is continued until all samples reach $\lambda_2$. We now have an estimate of $t_0^2$ and the locations along $\lambda_2$ where the sample paths ended. From these locations the paths are continued until they reach $\lambda_3$ or $\lambda_0$. If they reach $\lambda_0$ they are restarted at a location where a previous path stopped $\lambda_2$ and $t_0^2$ is added to the time. This step is finished once all sample paths reach $\lambda_3$. The general step is to start the paths on $\lambda_k$ and run them until they reach $\lambda_{k+1}$ or $\lambda_0$. If a path reaches $\lambda_0$, it is restarted at $\lambda_k$ with $t_0^k$ added to the time. The step is complete when all paths reach $\lambda_{k+1}$ and time time gives $t_0^{k+1}$. This process is repeated until $\lambda_M$ is reached. The result is a value for $t_0^M$ which is the estimate for the transition time.

This construction shows the relationship between FFS and the barrier method \cite{Adams2010Barrier}. This version of FFS measures the average time it takes to reach each barrier during the algorithm, as in the barrier method. Also both methods avoid characterizing the flux rate from the surface and the transition probability. The main difference between this algorithm and the barrier method is that the barrier method need the simulation to go all the way back to $\lambda_0$ before jumping back to the current barrier; only the previous barrier need be reached. This is the source of the performance gains of the barrier method. However, this method is currently only tractable for low-dimensional systems as it requires a reasonable sample of previous barriers.


\section{Appendix B --- Testing FFS and FFST on an exactly solvable problem}
\label{sec:AppendixB}

In this Appendix, we use a simple one-dimensional system to show that FFST can give accurate results for transition times when FFS fails. We also briefly discuss the comparative efficiency of the different algorithms on the ZGB model.

Consider a discrete hopping process on a line of length $L$. The probability of jumping from the $i^{\textrm{th}}$ to the $(i-1)^{\textrm{th}}$ site is $p_i$. The time of a jump is unity. In the cases where the first site ($i=0$) is adsorbing and the last site is reflecting ($p_{L-1} = 1$), the system can be solved exactly \cite{doering2005extinction}. The solution can be written in terms of hopping rates instead of hopping probability. $\lambda_i$ and $\mu_i$ are the rates of hopping from $i$ to $(i+1)$ and $(i-1)$ respectively. In terms of $p_i$ these are: $\mu_i = p_i$ and $\lambda_i = 1 - p_i$. The average time to reach the adsorbing site $i=0$ from site $i=n$ is \cite{doering2005extinction}:
\begin{equation}
\label{eq:ExactResult}
\begin{split}
\tau_n &=  \sum_{m=1}^n \left[ \frac{1}{\mu_m} + \prod_{i=1}^{m-1} \frac{\mu_i}{\lambda_i} \sum_{j=m+1}^{L-1} \frac{1}{\mu_j} \prod_{k=1}^{j-1} \frac{\lambda_k}{\mu_k} \right], \\
&= \sum_{m=1}^n \left[ \frac{1}{p_m} + \prod_{i=1}^{m-1} \frac{p_i}{1 - p_i} \sum_{j=m+1}^{L-1} \frac{1}{p_j} \prod_{k=1}^{j-1} \left( \frac{1}{p_k} -1  \right) \right].
\end{split}
\end{equation}
This equation has been used to find the extinction time of a disease within a population in a simple model from epidemiology \cite{doering2005extinction}. Equation (\ref{eq:ExactResult}) is general and can be used to construct `energy landscapes.' We create a landscape with non-uniform hopping probabilities such that there are three metastable states, regions $A$, $B$, and $C$, all with roughly equal stability, as shown in Fig.~\ref{fig:Accuracy}a. We measure the time it takes to reach the absorbing state starting near the reflecting boundary. This requires escape from the first metastable region $A$, then the second metastable region $B$, to finally reach the absorbing state near the center of the last metastable region $C$. We measured the time using FFS and FFST for various sizes of systems (well depths) as shown in Fig.~\ref{fig:Accuracy}b.  We found that FFS significantly underestimates the transition time by as much as $50$\%. There is also a significant increase in the variance of the result, roughly an order of magnitude for this model system. This is caused by the flux being strongly influenced by the rare occurrence of a trajectory that makes it to region $B$, spends a long time there, and then returns to region $A$.

Even in the absence of long-lived metastable states, FFST can produce transition times with less variance than FFS because it samples the external return time significantly better. We found this to be the case in the ZGB model. Using $L=32$, eleven evenly spaced static barriers starting at $C=0.06$, $N=10^3$, and $p_\mathrm{CO}=0.5268$, we found a variance of $5.0$\% for FFST and $5.8$\% for FFS; an improvement of about $16$\%. To equal the variance of the FFST result, FFS which would require roughly $35$\% more trials which would translate into a $35$\% longer run time.  In the case of the ZGB model, this effect can be mitigated by choosing $\lambda_0$ to be far enough from the metastable region that the internal return time $T_{int}$ is much larger than the external return time $T_{ext}$, bounding the effect on the increase in variance of $T_{ext}$. In general, moving $\lambda_0$ is not always useful, because metastable states can make $T_{ext}$ arbitrarily large.

\bibliographystyle{unsrt}


\begin{thebibliography}{10}

\bibitem{andersson2000stochastic}
H.~Andersson and T.~Britton.
\newblock {\em {Stochastic Epidemic Models and their Statistical Analysis}}.
\newblock Springer Verlag, 2000.

\bibitem{bartlett1961stochastic}
M.~S. Bartlett.
\newblock {\em {Stochastic Population Models in Ecology and Epidemiology}}.
\newblock Wiley, New York, 1961.

\bibitem{medhi2003stochastic}
J.~Medhi.
\newblock {\em {Stochastic Models in Queueing Theory}}.
\newblock Academic Press, Boston, 2003.

\bibitem{van2007stochastic}
N.~G. van Kampen.
\newblock {\em {Stochastic Processes in Physics and Chemistry}}.
\newblock North-Holland, 2007.

\bibitem{bolhuis2002t}
P.~G. Bolhuis, D.~Chandler, C.~Dellago, and P.~L. Geissler.
\newblock {Transition Path Sampling: Throwing Ropes Over Rough Mountain Passes,
  in the Dark}.
\newblock {\em Annu. Rev. Phys, Chem.}, 53(1):291--318, 2002.

\bibitem{van2003novel}
T.~S. van Erp, D.~Moroni, and P.~G. Bolhuis.
\newblock {A novel path sampling method for the calculation of rate constants}.
\newblock {\em J. Chem. Phys.}, 118:7762, 2003.

\bibitem{faradjian2004computing}
A.~K. Faradjian and R.~Elber.
\newblock {Computing time scales from reaction coordinates by milestoning}.
\newblock {\em J. Chem. Phys.}, 120:10880, 2004.

\bibitem{W2005Finite}
W.~E, W.~Q. Ren, and E.~Vanden-Eijnden.
\newblock Finite temperature string method for the study of rare events.
\newblock {\em J. Phys. Chem. B}, 109(14):6688, 2005.

\bibitem{Huber1996Weighted}
G.~A. Huber and S.~Kim.
\newblock Weighted ensemble brownian dynamics simulations for protein
  association reactions.
\newblock {\em Biophys. J.}, 70(1):97--110, 1996.

\bibitem{dellago2008transition}
C.~Dellago and P.~G. Bolhuis.
\newblock {Transition Path Sampling and other Advanced Simulation Techniques
  for Rare Events}.
\newblock {\em Advanced Computer Simulation Approaches for Soft Matter Sciences
  III}, 221:167--233, 2008.

\bibitem{allen2005sampling}
R.~J. Allen, P.~B. Warren, and P.~R. ten Wolde.
\newblock {Sampling rare switching events in biochemical networks}.
\newblock {\em Phys. Rev. Lett.}, 94(1):18104, 2005.

\bibitem{allen2006simulating}
R.~J. Allen, D.~Frenkel, and P.~R. ten Wolde.
\newblock {Simulating rare events in equilibrium or nonequilibrium stochastic
  systems}.
\newblock {\em J. Chem. Phys.}, 124:024102, 2006.

\bibitem{valeriani2007computing}
C.~Valeriani, R.~J. Allen, M.~J. Morelli, D.~Frenkel, and P.~R. ten Wolde.
\newblock {Computing stationary distributions in equilibrium and nonequilibrium
  systems with forward flux sampling}.
\newblock {\em J. Chem. Phys.}, 127:114109, 2007.

\bibitem{sanz2007evidence}
E.~Sanz, C.~Valeriani, D.~Frenkel, and M.~Dijkstra.
\newblock {Evidence for out-of-equilibrium crystal nucleation in suspensions of
  oppositely charged colloids}.
\newblock {\em Phys. Rev. Lett.}, 99(5):55501, 2007.

\bibitem{allen2008homogeneous}
R.~J. Allen, C.~Valeriani, S.~T{\u{a}}nase-Nicola, P.~R. ten Wolde, and
  D.~Frenkel.
\newblock {Homogeneous nucleation under shear in a two-dimensional Ising model:
  Cluster growth, coalescence, and breakup}.
\newblock {\em J. Chem. Phys.}, 129:134704, 2008.

\bibitem{Velez2009Kinetics}
C.~Velez-Vega, E.~Borrero, and F.~Escobedo.
\newblock Kinetics and reaction coordinate for the isomerization of alanine
  dipeptide by a forward flux sampling protocol.
\newblock {\em J. Chem. Phys.}, 130(22):225101, 2009.

\bibitem{maier1993effect}
R.~S. Maier and D.~L. Stein.
\newblock {Effect of focusing and caustics on exit phenomena in systems lacking
  detailed balance}.
\newblock {\em Phys. Rev. Lett.}, 71(12):1783--1786, 1993.

\bibitem{allen2006forward}
R.~J. Allen, D.~Frenkel, and P.~R. ten Wolde.
\newblock {Forward flux sampling-type schemes for simulating rare events:
  Efficiency analysis}.
\newblock {\em J. Chem. Phys.}, 124:194111, 2006.

\bibitem{ZiffGulariBarshad86}
R.~M. Ziff, E.~Gulari, and Y.~Barshad.
\newblock {Kinetic phase transitions in an irreversible surface-reaction
  model}.
\newblock {\em Phys. Rev. Lett.}, 56(24):2553--2556, 1986.

\bibitem{EhsasiEtAl89}
M.~Ehsasi, M.~Matloch, O.~Frank, J.~H. Block, K.~Christmann, F.~S. Rys, and
  W.~Hirschwald.
\newblock {Steady and nonsteady rates of reaction in a heterogeneously
  catalyzed reaction: Oxidation of CO on platinum, experiments and
  simulations}.
\newblock {\em J. Chem. Phys.}, 91:4949, 1989.

\bibitem{MachadoBuendiaRikvold05}
E.~Machado, G.~M. Buend{\'\i}a, and P.~A. Rikvold.
\newblock {Decay of metastable phases in a model for the catalytic oxidation of
  CO}.
\newblock {\em Physical Review E}, 71(3):31603, 2005.

\bibitem{MachadoBueniaRikvoldZiff05}
E.~Machado, G.~M. Buend{\'\i}a, P.~A. Rikvold, and R.~M. Ziff.
\newblock {Response of a catalytic reaction to periodic variation of the CO
  pressure: Increased CO\_ $\{$2$\}$ production and dynamic phase transition}.
\newblock {\em Phys. Rev. E}, 71(1):16120, 2005.

\bibitem{Jakubith..Ertl90}
S.~Jakubith, H.~H. Rotermund, W.~Engel, A.~von Oertzen, and G.~Ertl.
\newblock Spatiotemporal concentration patterns in a surface reaction:
  Propagating and standing waves, rotating spirals, and turbulence.
\newblock {\em Phys. Rev. Lett.}, 65(24):3013--3016, Dec 1990.

\bibitem{Kim..Ertl01}
M.~Kim, M.~Bertram, M.~Pollmann, A.~von Oertzen, A.~S. Mikhailov, H.~H.
  Rotermund, and G.~Ertl.
\newblock {Controlling Chemical Turbulence by Global Delayed Feedback: Pattern
  Formation in Catalytic CO Oxidation on Pt(110)}.
\newblock {\em Science}, 292(5520):1357--1360, 1998.

\bibitem{MarroDickman99}
J.~Marro and R.~Dickman.
\newblock {\em {Nonequilibrium Phase Transitions}}.
\newblock Cambridge University Press, 1999.

\bibitem{MeakinScalapino87}
P.~Meakin and D.~J. Scalapino.
\newblock Simple models for heterogeneous catalysis: Phase transition-like
  behavior in nonequilibrium systems.
\newblock {\em J. Chem. Phys.}, 87(1):731--741, 1987.

\bibitem{BrosilowZiff92}
B.~J. Brosilow and R.~M. Ziff.
\newblock Effects of a desorption on the first-order transition in the a-b$_2$
  reaction model.
\newblock {\em Phys. Rev. A}, 46(8):4534--4538, Oct 1992.

\bibitem{LoscarAlbano09}
E.~S. Loscar and E.~V. Albano.
\newblock Numerical study of the evaporation/condensation phase transition of
  droplets for an irreversible reaction model.
\newblock {\em EPL (Europhysics Letters)}, 85(3):30004, 2009.

\bibitem{ZiffBrosilow92}
R.~M. Ziff and B.~J. Brosilow.
\newblock {Investigation of the first-order phase transition in the A-B$_2$
  reaction model using a constant-coverage kinetic ensemble}.
\newblock {\em Phys. Rev. A}, 46(8):4630--4633, 1992.

\bibitem{TomeDickman93}
T.~Tom\'e and R.~Dickman.
\newblock Ziff-{G}ulari-{B}arshad model with {CO} desorption: An {I}sing-like
  nonequilibrium critical point.
\newblock {\em Phys. Rev. E}, 47(2):948--952, Feb 1993.

\bibitem{EvansMiesch91}
J.~W. Evans and M.~S. Miesch.
\newblock {Characterizing kinetics near a first-order catalytic-poisoning
  transition}.
\newblock {\em Phys. Rev. Lett.}, 66(6):833--836, 1991.

\bibitem{EvansRay94}
J.~W. Evans and T.~R. Ray.
\newblock Interface propagation and nucleation phenomena for discontinuous
  poisoning transitions in surface-reaction models.
\newblock {\em Phys. Rev. E}, 50(6):4302--4314, 1994.

\bibitem{GoodmanEtAl95}
R.~H. Goodman, D.~S. Graff, L.~M. Sander, P.~Leroux-Hugon, and E.~Cl\'ement.
\newblock Trigger waves in a model for catalysis.
\newblock {\em Phys. Rev. E}, 52(6):5904--5909, 1995.

\bibitem{Albano01}
E.~V. Albano.
\newblock Monte {C}arlo simulations of the short time dynamics of a first-order
  irreversible phase transition.
\newblock {\em Physics Letters A}, 288(2):73 -- 78, 2001.

\bibitem{Borrero08}
E.~Borrero and F.~Escobedo.
\newblock Optimizing the sampling and staging for simulations of rare events
  via forward flux sampling schemes.
\newblock {\em J. Chem. Phys.}, 129(2):024115, 2008.

\bibitem{Adams2010Barrier}
D.~A. Adams, L.~M. Sander, and R.~M. Ziff.
\newblock The barrier method: A technique for calculating very long transition
  times.
\newblock {\em arXiv:1005.3985}, 2010.

\bibitem{weinan2003energy}
W.~E, W.~Ren, and E.~Vanden-Eijnden.
\newblock {Energy landscape and thermally activated switching of
  submicron-sized ferromagnetic elements}.
\newblock {\em Journal of Applied Physics}, 93:2275, 2003.

\bibitem{catlett2007teragrid}
L.~Grandinetti, editor.
\newblock {\em {High Performance Computing and Grids in Action}}.
\newblock IOS Press, Amsterdam, 2008.

\bibitem{allen2009forward}
R.~J. Allen, C.~Valeriani, and P.~R. ten Wolde.
\newblock {Forward flux sampling for rare event simulations}.
\newblock {\em J. Phys: Cond. Matter}, 21:463102, 2009.

\bibitem{doering2005extinction}
C.~R. Doering, K.~V. Sargsyan, and L.~M. Sander.
\newblock {Extinction Times for Birth-Death Processes: Exact Results, Continuum
  Asymptotics, and the Failure of the Fokker--Planck Approximation}.
\newblock {\em Multiscale Model. Simul.}, 3(2), 2005.

\end{thebibliography}

\end{document}